\documentclass[pdflatex,sn-mathphys-num]{sn-jnl}

\usepackage{graphicx}%
\usepackage{multirow}%
\usepackage{amsmath,amssymb,amsfonts}%
\usepackage{amsthm}%
\usepackage{mathrsfs}%
\usepackage[title]{appendix}%
\usepackage{xcolor}%
\usepackage{textcomp}%
\usepackage{manyfoot}%
\usepackage{booktabs}%
\usepackage{algorithm}%
\usepackage{algorithmicx}%
\usepackage{algpseudocode}%
\usepackage{listings}%
\usepackage{tikz}
\usepackage{forest}
\usepackage{tikz-qtree}
\usepackage{array}
\usepackage{subcaption}

\theoremstyle{thmstyleone}%
\theoremstyle{thmstyletwo}%

\theoremstyle{thmstylethree}%

\begin{document}

\title[A Hybrid TGN-SEAL Model for Dynamic Graph Link Prediction]{A Hybrid TGN-SEAL Model for Dynamic Graph Link Prediction}


\author[1]{Nafiseh Sadat Sajadi}
\author[2]{Behnam Bahrak}
\author[1]{Mahdi Jafari Siavoshani}

\affil[1]{Department of Computer Engineering, Sharif University of Technology, Tehran, Iran}
\affil[2]{Tehran Institute for Advanced Studies, Khatam University, Tehran, Iran}


\abstract{Predicting links in sparse, continuously evolving networks is a central challenge in network science. Conventional heuristic methods and deep learning models, including Graph Neural Networks (GNNs), are typically designed for static graphs and thus struggle to capture temporal dependencies. Snapshot-based techniques partially address this issue but often encounter data sparsity and class imbalance, particularly in networks with transient interactions such as telecommunication call detail records (CDRs). Temporal Graph Networks (TGNs) model dynamic graphs by updating node embeddings over time; however, their predictive accuracy under sparse conditions remains limited. In this study, we improve the TGN framework by extracting enclosing subgraphs around candidate links, enabling the model to jointly learn structural and temporal information. Experiments on a sparse CDR, email, message dataset show that our approach increases average precision by at least $2\%$ over standard TGNs, demonstrating the advantages of integrating local topology for robust link prediction in dynamic networks.}


\maketitle


\section{Introduction}

Modeling complex systems as graphs provides a unifying framework for diverse domains, from social interactions to biological networks, where nodes are entities and edges represent the relationships between them~\cite{hamilton2020graph}. As the volume of graph-structured data has grown, research efforts have increasingly focused on learning methods that effectively use this structure for common tasks, including node classification, graph classification, and link prediction. In this study, we focus on predicting future links in dynamic graphs, where the goal is to forecast how relationship between nodes evolve as the network changes over time.

Many real-world networks change continuously, with nodes and edges appearing, disappearing, or being updated as events occur. In this work, we study this problem in the context of a user call network, aiming to predict future communication events based on historical call detail records that include information such as caller, recipient, timestamp, and call duration. 

Dynamic link prediction has broad practical value across numerous fields. In telecommunications, for example, forecasting future call and interaction patterns allows operators to allocate resources more efficiently, suggest customized services based on user needs, and optimize staffing schedules~\cite{jalal2016forecasting}. Beyond this domain, similar techniques are essential in areas like e-commerce (e.g., recommender systems,) bioinformatics (e.g., predicting protein–protein interactions and drug discovery), and various security applications~\cite{link_pred_social_networks}.

To address link prediction in sparse dynamic networks, we introduce TGN-SEAL, a hybrid framework that enhances Temporal Graph Networks by extracting and leveraging structural information from the enclosing subgraph for each candidate link. By capturing both temporal patterns and localized topology, our approach improves link prediction performance in sparse, continuously evolving networks.


\section{Related Work}

In link prediction, the objective is to estimate the probability that a connection will form between two given nodes in a graph. Owing to the pervasive use of network structures, this problem has a wide range of applications, including in recommender systems, knowledge graph completion, and the reconstruction of metabolic networks.

A widely used and effective family of straightforward approaches for link prediction is heuristic methods. These methods assign a score to each pair of nodes, representing the likelihood that a link will form between them. Heuristics are typically classified based on the number of steps required to reach neighboring nodes. For example, common neighbors and preferential attachment are first-order heuristics, as they consider direct neighbors, while Adamic–Adar is a second-order method because it relies on information from nodes two steps away~\cite{adamic2003friends}.

In general, a heuristic of order $h$ requires information about the $h$-hop neighborhood of the target nodes. Some advanced heuristics, such as Katz~\cite{katz1953new}, rooted PageRank~\cite{haveliwala2002topic}, and SimRank~\cite{jeh2002simrank}, go further by leveraging global graph information. Despite their efficiency and success in some contexts, these methods often rely on rigid assumptions about how links form. For instance, the common neighbors heuristic assumes that nodes sharing more neighbors are more likely to connect. While this assumption holds true in social networks, it does not apply to all types of networks. In protein–protein interaction networks, for example, the likelihood of interaction tends to increase when nodes share fewer neighbors, demonstrating how the same heuristic may fail in different domains.

Liben-Nowell and Kleinberg~\cite{link_pred_liben} introduced one of the earliest models for link prediction in social networks. They represented social interactions using a graph, where nodes corresponded to individuals and edges captured the relationships between them. The number of interactions between two person was modeled using either multigraph or weighted graph. Their approach was unsupervised, focusing on evaluating node similarity based on similarity measures, followed by ranking potential links according to these scores. They demonstrated that similarity-based heuristics such as common neighbors, preferential attachment, Adamic–Adar, Jaccard, SimRank, hitting time, rooted PageRank, and Katz could outperform random link prediction~\cite{haghani2019systemic}. Subsequently, Hasan et al.~\cite{link_pred_suprvised} expanded this work in two key ways: first, by emphasizing the importance of incorporating social network data alongside graph topology, which substantially improved predictive performance, and second, by reformulating the problem as a supervised binary classification task that uses similarity indices as input features to predict the presence or absence of a link between pairs of nodes~\cite{haghani2019systemic}.

Lü and Zhou~\cite{lu2011link} later proposed a comprehensive taxonomy of link prediction methods, classifying them into three main categories: (1) similarity-based approaches, (2) maximum likelihood methods, and (3) probabilistic models. They reported that a significant drawback of maximum likelihood methods is their computational inefficiency, while probabilistic models are designed to optimize a built target function to best fit the observed data. Furthermore, they emphasized studies on these models conducted by statistical physicists~\cite{haghani2019systemic}. Notable examples of these models include Probabilistic Relational Models (PRMs)~\cite{getoor2002learning}, Graphical Models~\cite{nallapati2008joint}, and Stochastic Relational Models~\cite{airoldi2006mixed}, which are mainly developed for homogeneous networks.. Although these models can handle heterogeneous data, they often introduce considerable complexity and encounter interpretability challenges~\cite{link_pred_social_networks}.

With the emergence of large-scale dynamic networks, traditional methods face challenges in scalability and adaptability. Consequently, machine learning and deep learning approaches have become increasingly prominent. These models treat graphs as non-Euclidean data structures and are employed in tasks such as node classification, clustering, and link prediction. Graph neural networks have attracted significant attention due to their strong predictive performance and interpretability~\cite{introduction_to_GNN_liu}.

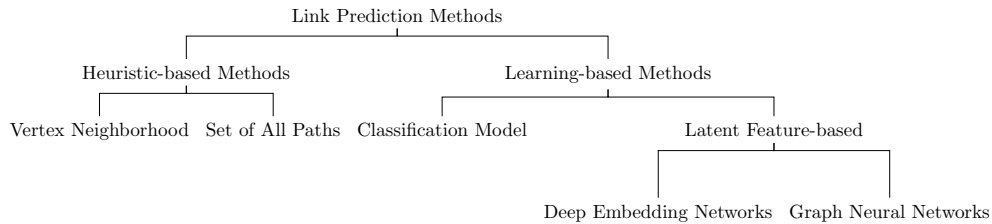
\begin{figure*}
\centering
\begin{tikzpicture}[
	scale=0.72,
	level 1/.style={level distance=1.7cm}
	level 2/.style={level distance=1.6cm},
    level 3/.style={level distance=1.5cm},
    every node/.style={align=center, },
    edge from parent/.style={draw,
						  edge from parent path={(\tikzparentnode.south)
											-- +(0,-4pt)
											-| (\tikzchildnode)}}
]
\Tree
[.{Link Prediction Methods}
	[.{Heuristic-based Methods}
		[.{Vertex Neighborhood} ]
		[.{Set of All Paths} ]]
         [.{Learning-based Methods}
         	[.{Classification Model} ] 
                 [.{Latent Feature-based}
                 	[.{Deep Embedding Networks} ]
                 	[.{Graph Neural Networks} ]]]
]
\end{tikzpicture}
\caption{Classification of Link Prediction Methods}
\label{link-pred-approach}
\end{figure*}

Given the importance of this problem and its numerous applications across various domains, researchers have continually tried to develop different methods, often guided by the specific problem domain they were addressing. As illustrated in Figure~\ref{link-pred-approach}, link prediction strategies can be divided into two major families: (1) heuristic-based and, (2) learning-based approaches~\cite{haghani2019systemic}.

\subsection{Heuristic-Based Methods}

Heuristic approaches estimate link likelihood using similarity scores derived from shared node characteristics or graph topology. They are commonly divided into node-based and path-based methods and, despite their simplicity and lack of temporal modeling, consistently outperform random baselines~\cite{lu2011link}. In practice, candidate node pairs are ranked by their scores and the top $k$ pairs are selected as predicted links. Because these methods capture tendencies such as homophily, the preference for nodes to connect to similar nodes, they are also referred to as similarity-based methods~\cite{nickel2015review, haghani2019systemic}.

\subsection{Learning-Based Methods}

Learning-based methods infer predictive patterns directly from data, capturing structural dependencies beyond handcrafted heuristics. They are broadly categorized into classification-based models and latent feature models~\cite{haghani2019systemic, dunlavy2011temporal}. The main distinction lies in the input representation: classification models rely on engineered features (e.g., similarity indices or node attributes), whereas latent feature models learn representations directly from the graph structure.

\subsubsection{Classification Models}

Classification approaches formulate link prediction as a supervised binary task, distinguishing observed (positive) from unobserved (negative) links~\cite{lee2013link}. Common classifiers include decision trees and support vector machines, with SVMs often achieving strong performance~\cite{nguyen2015transfer}. Unlike heuristic methods, these models learn from data and can generalize beyond predefined similarity measures~\cite{link_pred_suprvised}.

\subsubsection{Latent Feature-Based Models}

Latent feature models assume that graphs contain hidden structural patterns not directly observable from topology~\cite{sewell2016latent}. They learn node representations that capture complex relational dependencies, making them suitable for large or heterogeneous networks~\cite{haghani2019systemic}. Two prominent categories are:

\begin{enumerate}
\item \textbf{Embedding-Based Methods:} Inspired by natural language processing, approaches such as node embeddings learn low-dimensional representations that preserve structural and neighborhood information~\cite{grover2016node2vec, socher2013reasoning}.

\item \textbf{Graph Neural Networks:} GNNs extend neural models to graph data by iteratively aggregating information from neighbors through message passing, enabling joint modeling of topology and node attributes~\cite{zhou2018graph}.

\end{enumerate}


\section{Background}

Given the effectiveness and robustness of GNNs in various link prediction tasks, we focus on two GNNs models: Subgraph Embeddings and Attributes for Link prediction (SEAL) and Temporal Graph Networks (TGN). SEAL model is designed for static graphs, leveraging subgraph extraction and structure-aware learning. In contrast, TGN is developed for dynamic graphs, incorporating temporal message-passing mechanisms to capture evolving node representations. These models represent modern neural link prediction techniques for both static and temporal networks.

\subsection{Subgraph Embeddings and Attributes for Link Prediction}
As discussed earlier, heuristic-based methods such as common neighbors or the Katz index estimate link probabilities using predefined scoring functions. While these methods are simple, interpretable, and scalable, they rely on rigid assumptions about link formation, which limits their effectiveness. A more flexible approach is to learn heuristics directly from data.

M. Zhang et al.~\cite{zhang2018link} proposed SEAL, a framework that extracts a local subgraph around each target link and employs a graph neural network to learn structural representations. The GNN is trained as a binary classifier on these subgraphs, learning to identify structural patterns that predict link formation. Unlike traditional heuristics, SEAL does not depend on hand-crafted similarity measures but instead learns general structural features directly from the data.

To formalize this, Zhang et al. introduced the unified $\gamma$-decaying theory, showing that many higher-order heuristics can be approximated using information from enclosing subgraphs. They encoded each subgraph's structure into a node information matrix, which the GNN uses to learn topological features for prediction task. They introduced the Weisfeiler-Lehman Neural Machine (WLNM), a fully connected neural architecture for link prediction. WLNM takes enclosing subgraphs as input, which are induced by the $h$-hop neighbors of the target nodes $(x, y)$. These subgraphs capture relevant topological information; for instance, the number of common neighbors can be directly computed from a $1$-hop enclosing subgraph. The method performed competitively on several standard link prediction benchmarks~\cite{zhang2018link}.

\subsection{Temporal Graph Network}
Conventional graph neural networks are designed for static graphs and therefore cannot explicitly model temporal dependencies in evolving networks. However, many real-world systems such as social or communication networks are inherently dynamic, where edges and nodes appear continuously over time. While static GNNs can be applied by ignoring temporal evolution, this simplification often results in suboptimal performance, as temporal information frequently encodes critical insights about the system’s behavior. To address these limitations, the Temporal Graph Network was introduced as a general framework for learning on continuous-time dynamic graphs. Instead of representing a network as a sequence of discrete snapshots, TGN models its evolution as a stream of time-stamped interaction events. It combines memory modules with message-passing mechanisms to continuously track and update node representations as the graph evolves.

Dynamic graphs can be categorized based on how they represent time. Discrete-time dynamic graphs (DTDGs) model temporal evolution as a sequence of static snapshots captured at fixed intervals. In contrast, continuous-time dynamic graphs (CTDGs) represent the network as a continuous stream of time-stamped events such as the addition, deletion, or modification of nodes and edges. The Temporal Graph Network is specifically designed for CTDGs, viewing the graph as an ordered sequence of events $G = {x(t_1), x(t_2), \ldots}$, where each $x(t)$ corresponds to an interaction between one or more nodes at time $t$.

Following the terminology of Kazemi et al.~\cite{kazemi2020representation}, a neural model for dynamic graphs can be viewed as an encoder–decoder framework. The encoder generates time-dependent node embeddings $\mathbf{Z}(t)$, while the decoder performs downstream tasks such as link prediction or node classification. A central component of this architecture is the memory vector $\mathbf{s}_i(t)$ for each node $i$, which is updated only when relevant events occur. This memory acts as a compact summary of the node’s historical interactions, enabling the model to maintain temporal context over time.

When a pairwise interaction event $\mathbf{e}_{ij}(t)$ occurs, the model computes messages for both participating nodes using their previous memory states $\mathbf{s}_i(t^-)$ and $\mathbf{s}_j(t^-)$, the elapsed time $\Delta t$, and the associated event features. In the case of a node-wise event $\mathbf{v}_i(t)$, a single message is generated based on the node’s current memory and event attributes. If multiple messages are associated with the same node within a batch, they are aggregated into a single representation $\bar{\mathbf{m}}_i(t)$ using an aggregation function such as the mean or most-recent operation. The node’s memory is then updated according to $\mathbf{s}_i(t) = \text{mem}(\bar{\mathbf{m}}_i(t), \mathbf{s}_i(t^-))$, allowing both pairwise and individual events to contribute to the evolving state of the node.

Updated node embeddings $\mathbf{z}_i(t)$ are dynamically computed by integrating the node’s current memory, the memories of its temporal neighbors, and features from recent events. This mechanism reduces memory staleness and ensures that node representations remain up to date with ongoing interactions. During training, events are processed sequentially in temporal order to preserve causal consistency and avoid information leakage from future events.

Overall, the Temporal Graph Network provides a general and efficient framework for learning on continuous-time dynamic graphs. It achieves state-of-the-art performance while maintaining computational efficiency. Ablation studies further demonstrate that both the memory component and the graph-based embedding module are essential to the model’s effectiveness~\cite{rossi2020temporal}.


\section{Methodology}

We propose a framework for continuous-time link prediction in dynamic networks. Building on the Temporal Graph Network architecture, our approach leverages its memory modules and temporal embedding mechanisms as a foundation. In the TGN framework, each node maintains a dedicated memory state that is updated upon the occurrence of an event, enabling the model to process interactions continuously as they unfold over time.

\begin{figure*}[htbp]
\centering
\vspace{1em}
\includegraphics[width=0.7\textwidth]{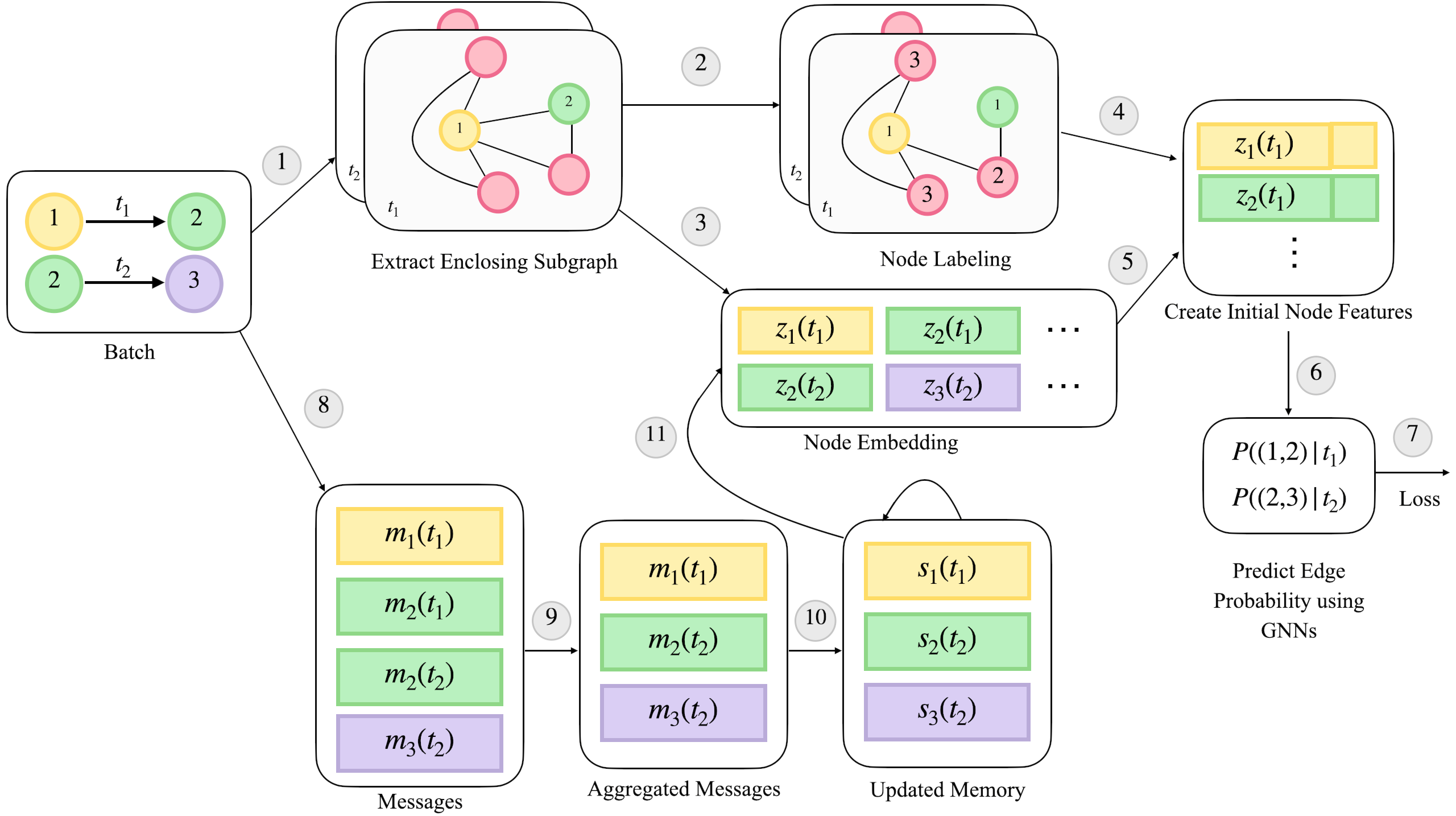}
\vspace{1em}
\caption{Proposed architecture of the TGN-SEAL framework.}
\label{fig:tgn-seal-architecture}
\end{figure*}

\subsection{Proposed Architecture: TGN-SEAL}

In the standard TGN setup, the temporal embeddings of two target nodes are concatenated and passed through a multi-layer perceptron (MLP) to predict the likelihood of a link. We enhance this link prediction module by replacing it with a new subgraph-based strategy. Our approach builds inspired by SEAL framework~\cite{zhang2018link}, which demonstrated that high-order heuristics can be approximated by extracting local neighborhoods. We therefore replace the TGN's prediction module with one that extracts and processes these local enclosing subgraphs. More precisely, at each iteration, the TGN-SEAL model first computes the temporal embedding for each node and these embeddings serve as initial feature vectors. They are then concatenated with other features, such as the topological position of each node, and the complete set of features is fed into a graph neural network for prediction. This redefines the link prediction problem as a graph classification task. Instead of predicting a link based only on the two target nodes, the GNN makes a decision based on the entire extracted subgraph. The proposed architecture is shown in Figure~\ref{fig:tgn-seal-architecture}.

\subsection{Training Procedure}

The training process for each batch of temporal events is performed as follows:

\begin{enumerate}
	\item
	\textbf{Subgraph Extraction}: For each interaction between a source node $u$ and a destination node $v$ at time $t$, we extract an enclosing subgraph. This subgraph is induced by the $k$-hop neighborhood around $u$ and $v$ and includes only the interactions that occurred prior to time $t$. Following previous work, we set $k=2$ and $k=3$ to capture higher-order structural patterns.

	\item
	\textbf{Structural Labeling}: Each node within the extracted subgraph is assigned a structural label using the Double-Radius Node Labeling (DRNL) method, which uses the pair of shortest path distances to the target nodes, to capture its topological role relative to the target nodes $(u,v)$. The DRNL scheme encodes the distance of each node to the two target nodes in a closed-form manner, allowing efficient computation even for large subgraphs. These labels are then converted into one-hot vectors and used as input features for the subgraph-based link prediction model.~\cite{zhang2018link}.

	\item
	\textbf{Temporal Representation}: We compute the temporal embedding for every node in the subgraph using the TGN's embedding module (e.g., the identity module). This embedding is a function of the node's memory state, summarizing its recent activity.

	\item
	\textbf{Feature Concatenation}: The final feature vector for each node is constructed by concatenating its temporal representation (from Step 3) with its one-hot encoded structural label (from Step 2).

	\item
	\textbf{Link Probability Prediction}: The enclosing subgraph, along with the final node feature vectors, is input into a graph neural network. In this study, we use the DGCNN architecture~\cite{dgcnn}, which outputs a scalar prediction $\hat{y} \in [0, 1]$ representing the probability of a link between $u$ and $v$. The prediction error is then calculated using binary cross-entropy loss.

	\item
	\textbf{Message Generation}: After computing link predictions for all candidate node pairs in the current batch using their temporal enclosing subgraphs, the memory update phase is initiated. For each interaction $(u,v,t)$ occurring in the batch, messages associated with the source and destination nodes are generated.
Each message is constructed from the memory states of the corresponding nodes prior to processing the batch, supplemented by their current embeddings obtained from the TGN-SEAL module, the features of the interaction (if available), and a temporal encoding of the interaction time. In the proposed TGN-SEAL framework, the embeddings used in message construction incorporate structural information derived from the $k$-hop enclosing subgraphs; however, the message generation mechanism itself follows the standard TGN formulation.

	\item
	\textbf{Message Aggregation}: Since a node may participate in multiple interactions within the same batch, multiple messages can be produced for a single node. To obtain a unique update signal per node, these messages must be aggregated.
In this work, we adopt the last-message aggregation strategy for computational efficiency and consistency with the original TGN design. Specifically, for each node, only the most recent message in temporal order within the batch is retained and used for updating its memory. This approach preserves the latest interaction information while avoiding the overhead associated with aggregating all messages.

	\item
	\textbf{Memory Update}: The aggregated messages are used to update the memory states of the corresponding nodes via the chosen memory updater. The updated memory is then stored and subsequently used to compute node embeddings for the next batch of interactions.

Importantly, memory updates are performed only after all predictions for the current batch have been completed. This delayed update mechanism ensures that the node representations used for predicting interactions at time $t$ depend exclusively on information available prior to $t$, thereby preventing information leakage from the target interactions. In the context of TGN-SEAL, this property is essential because enclosing subgraphs are extracted using only past interactions; postponing the memory update preserves the temporal causality and inductive validity of the learning process.

\end{enumerate}

\subsection{Datasets}
To evaluate the generality and robustness of the TGN-SEAL framework, we conducted experiments across a diverse range of temporal interaction networks. These datasets represent continuous-time dynamic graphs, modeled as sequences of timed events $G = \{x(t_1), x(t_2), ...\}$ where interactions can occur at any point in time. Each interaction is treated as a directed temporal edge in a multigraph, allowing for recurring interactions between the same pair of nodes.

The selected datasets span different social and communication domains, focusing on user-to-user interaction topologies:
\begin{enumerate}
\item {Reality Mining:} A telecommunication dataset capturing 87,417 mobile phone call events among 94 participants over a nine-month period. It is characterized by high structural sparsity and transient communication patterns~\cite{eagle2006reality}.
\item {Email-Eu-core:} Represents institutional email exchanges within a large European research institution. The network reflects structured communication across four distinct departmental subnetworks, totaling approximately 60,000 interactions~\cite{paranjape2017motifs}.
\item {CollegeMsg:} Records 59,835 private messages exchanged within an online university social network over a period of 193 days~\cite{panzarasa2009patterns}.
\end{enumerate}

\subsubsection{ Data Preprocessing and Features}
For all datasets, we adopted a $70\%–15\%–15\%$ chronological split for training, validation, and testing to preserve the temporal order of events and ensure that the model is tested on truly future interactions. In line with standard benchmarks for these networks, explicit node features are absent; therefore, all nodes were assigned a constant zero feature vector, forcing the model to rely exclusively on structural pattern and temporal history. For datasets without raw edge attributes (e.g., CollegeMsg), zero vectors were used as edge features for the message function. In contrast, for the CDR dataset, we utilized call duration as the edge feature, providing informative interaction strength signals to the model.

\begin{table*}[t]
\scriptsize
\centering
\caption{Summary statistics of the temporal datasets used in the experiments.}
\label{tab:dataset_stats}
\begin{tabular}{lrrrr}
\hline
Dataset & Nodes & Temporal Edges & Static Edges & Time Span (days) \\
\hline
Email-Eu-core Dept1 & 309   & 61,046 & 3,031  & 803 \\
Email-Eu-core Dept2 & 162   & 46,772 & 1,772  & 803 \\
Email-Eu-core Dept3 & 89    & 12,216 & 1,506  & 802 \\
Email-Eu-core Dept4 & 142   & 48,141 & 1,375  & 803 \\
\hline
CollegeMsg          & 1,899 & 59,835 & 20,296 & 193 \\
Reality Mining      & 94    & 87,417 & --     & $\sim$270 \\
\hline
\end{tabular}
\end{table*}


\section{Results}

We evaluated the proposed TGN-SEAL framework against the standard TGN model and several state-of-the-art temporal graph methods across diverse continuous-time interaction domains, including telecommunication, institutional email communication, online social interaction, and community question–answer platforms. These datasets exhibit varying communication patterns, temporal dynamics, and structural complexity, providing a comprehensive assessment of the framework’s ability to generalize across heterogeneous temporal networks.

\subsection{Performance Metrics and Baselines}

We report Mean Average Precision (mAP) as the primary evaluation metric, as it is well suited to dynamic link prediction tasks characterized by severe class imbalance.

The baseline models include three representative dynamic graph networks, JODIE, DyRep, and TGAT as well as three TGN variants used for ablation of its core components:

\begin{enumerate}
\item \textbf{TGN-no-mem:} Evaluates the necessity of the memory module for capturing long-term historical node states.
\item \textbf{TGN-id:} Assesses the impact of replacing the learnable message function with a simple identity mapping.
\item \textbf{TGN-time:} Isolates the contribution of temporal encoding to representation quality.
\end{enumerate}

In all experiments, we employed the \textit{last-message} aggregator and a GRU-based memory updater, which together provide a balance between computational efficiency and the ability to maintain up-to-date node representations.

\subsection{Comparative Analysis}

Tables~\ref{table:tgn-seal-result1}–\ref{table:tgn-seal-result3} report performance in both transductive and inductive settings. The results for TGN-SEAL correspond to the configuration with 2-hop enclosing subgraphs ($k=2$).

Overall, TGN-SEAL achieves the best or second-best results on most datasets, indicating that incorporating local structural subgraphs can substantially enhance temporal representation learning. The gains are particularly pronounced on datasets with stable structural patterns and recurring interactions, such as Reality Mining and the Email networks, where local topology provides strong predictive signals beyond temporal dynamics alone.

On the CollegeMsg dataset, TGN-SEAL ranks second, while TGAT and TGN-no-mem achieve the highest performance. This behavior suggests that link formation in this network is driven more by short-term temporal activity than by persistent structural patterns, making purely temporal or attention-based models particularly effective. The result underscores that the benefit of incorporating structural subgraphs depends on the characteristics of the underlying network.

The ablation study further highlights the complementary contributions of memory, temporal encoding, and structural context within the TGN framework. While baseline TGN variants already perform competitively, augmenting them with SEAL-based enclosing subgraphs yields consistent improvements on datasets where higher-order structural information is informative, demonstrating that structural and temporal signals play different roles depending on the domain.

\subsection{Convergence and Computational Efficiency}

\begin{figure*}[htbp]
\centering
\begin{subfigure}{0.32\textwidth}
    \includegraphics[width=\linewidth]{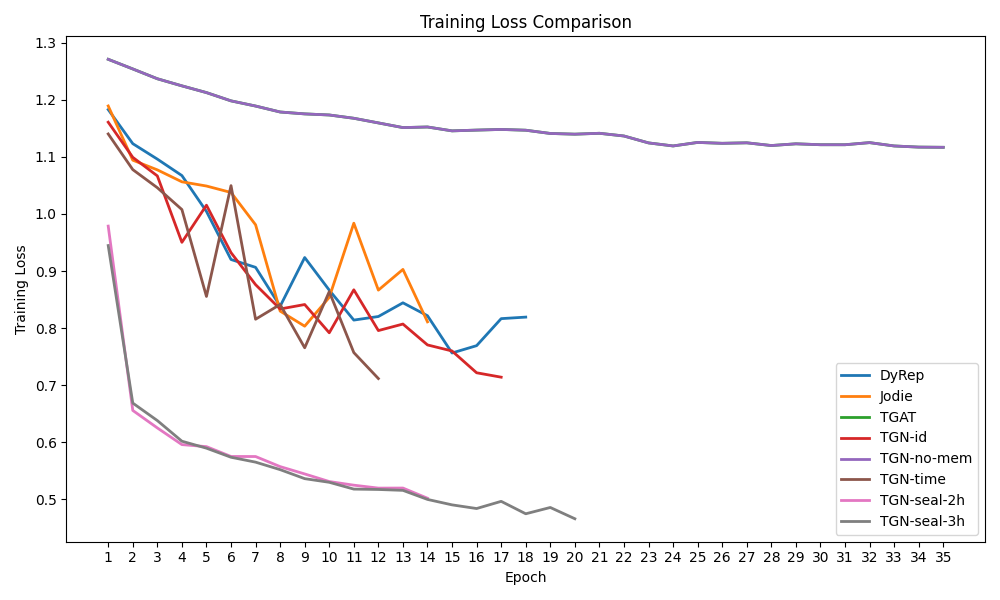}
    \caption{Training Loss}
    \label{subfig:train_loss}
\end{subfigure}
\hfill
\begin{subfigure}{0.32\textwidth}
    \includegraphics[width=\linewidth]{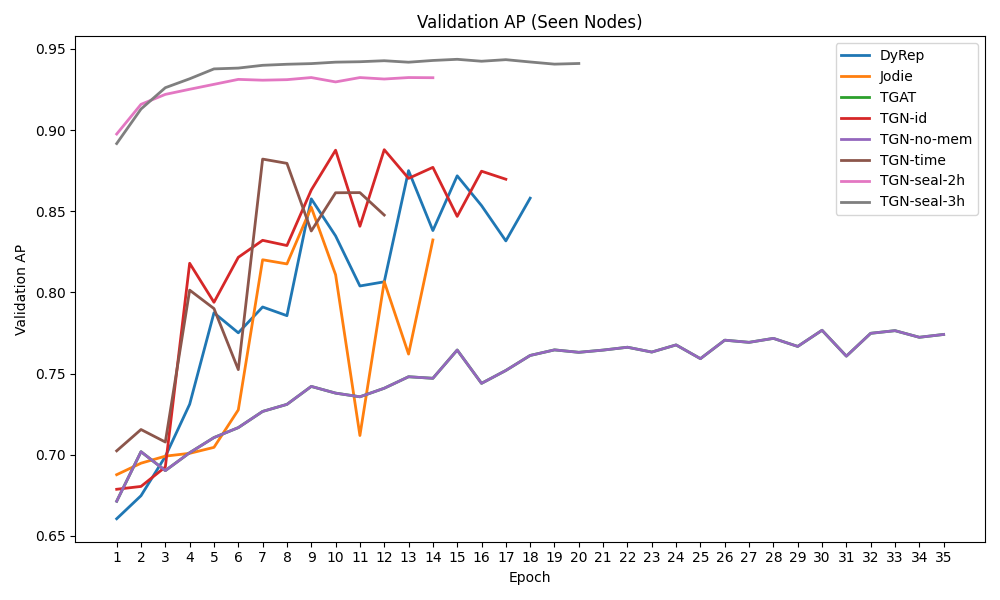}
    \caption{mAP on Seen Nodes}
    \label{subfig:seen_nodes}
\end{subfigure}
\hfill
\begin{subfigure}{0.32\textwidth}
    \includegraphics[width=\linewidth]{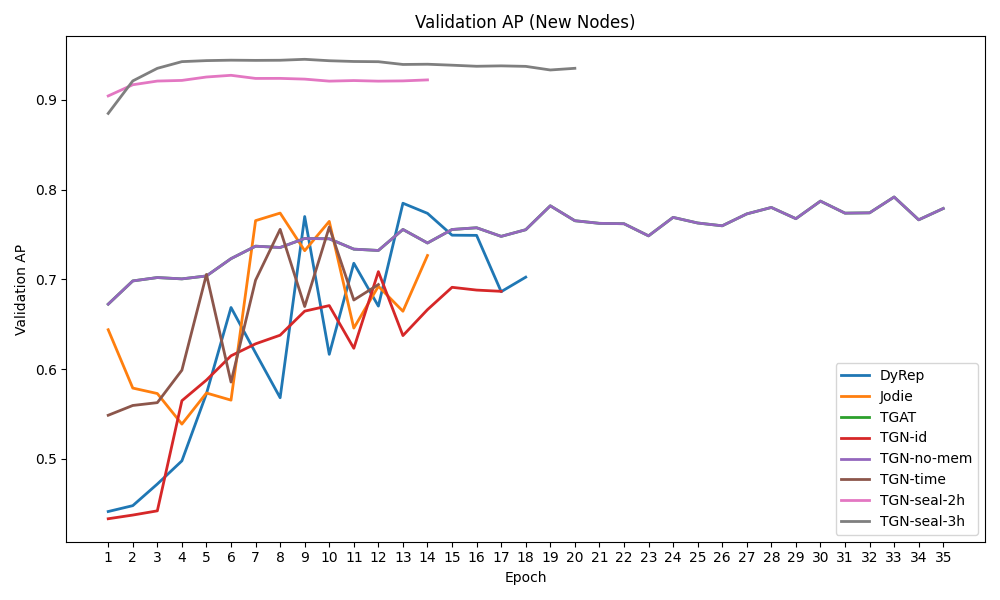}
    \caption{mAP on Unseen Nodes}
    \label{subfig:unseen_nodes}
\end{subfigure}

\caption{Training loss and mAP of TGN-SEAL and baseline methods on Email-Eu-core dataset. The framework achieves faster convergence and superior performance on both seen and unseen nodes.
TGN‑SEAL‑2h denotes the TGN‑SEAL model using a 2‑hop enclosing subgraph, while TGN‑SEAL‑3h denotes the variant using a 3‑hop enclosing subgraph.}
\label{fig:tgn-seal-all}
\end{figure*}

As shown in Figure~\ref{fig:tgn-seal-all}, TGN-SEAL converges in fewer training epochs than the standard TGN model, although each epoch is more computationally expensive due to subgraph extraction and aggregation. Increasing the neighborhood radius to $k=3$ may capture additional structural motifs, but it does not always lead to higher mAP, as the benefit depends on the graph topology. However, this setting increases per-epoch runtime by approximately $2.7\times$ compared to $k=2$, highlighting a trade-off between predictive gain and computational cost.

This behavior reflects a practical trade-off between predictive performance and computational cost. Depending on dataset size, sparsity, and domain requirements, practitioners can select the neighborhood radius that best balances accuracy and efficiency. The results suggest that $k=2$ already captures most of the informative local structure, while $k=3$ provides incremental gains when additional structural context is beneficial.

\subsection{Scalability Considerations}

To control computational overhead, the framework employs temporal neighbor sampling that prioritizes the most recent and informative interactions. This strategy preserves essential structural signals while limiting the size of extracted subgraphs, enabling the method to scale to large, evolving temporal networks. Combined with the localized nature of enclosing subgraphs, this design ensures that TGN-SEAL remains computationally feasible without sacrificing predictive performance.

\begin{table*}[t]
\centering
\caption{Performance comparison on Reality Mining and CollegeMsg datasets. 
\textbf{Bold}: best performance; \underline{Underline}: second best.}
\label{table:tgn-seal-result1}%
\resizebox{\textwidth}{!}{%
\begin{tabular}{lcccc}
\toprule
Method & \multicolumn{2}{c}{Reality Mining} & \multicolumn{2}{c}{CollegeMsg} \\
\cmidrule(lr){2-3} \cmidrule(lr){4-5}
& Transductive & Inductive & Transductive & Inductive \\
\midrule
DyRep
& $0.905 \pm 0.005$ & $0.945 \pm 0.005$
& $0.7946 \pm 0.0036$ & $0.6661 \pm 0.0196$ \\

JODIE
& $0.909 \pm 0.008$ & $0.949 \pm 0.007$
& $0.7878 \pm 0.0075$ & $0.6838 \pm 0.0199$ \\

TGAT
& $0.793 \pm 0.009$ & $0.824 \pm 0.012$
& $\pmb{0.8891 \pm 0.0022}$ & $\pmb{0.8770 \pm 0.0019}$ \\

TGN-no-mem
& $0.793 \pm 0.009$ & $0.824 \pm 0.012$
& $\pmb{0.8891 \pm 0.0022}$ & $\pmb{0.8770 \pm 0.0019}$ \\

TGN-id
& $\underline{0.919 \pm 0.006}$ & $0.960 \pm 0.004$
& $0.8068 \pm 0.0031$ & $0.7168 \pm 0.0169$ \\

TGN-time
& $0.917 \pm 0.006$ & $\underline{0.960 \pm 0.003}$
& $0.7984 \pm 0.0070$ & $0.6910 \pm 0.0229$ \\

\textbf{TGN-SEAL}
& $\pmb{0.945 \pm 0.002}$ & $\pmb{0.976 \pm 0.001}$
& $\underline{0.8090 \pm 0.0042}$ & $\underline{0.7822 \pm 0.0211}$ \\
\bottomrule
\end{tabular}%
}
\end{table*}

\begin{table*}[t]
\centering
\caption{Performance comparison on Email datasets (department 1 and 2).
\textbf{Bold}: best performance; \underline{Underline}: second best.}
\label{table:tgn-seal-result2}%
\resizebox{\textwidth}{!}{%
\begin{tabular}{lcccc}
\toprule
Method & \multicolumn{2}{c}{Email-1} & \multicolumn{2}{c}{Email-2} \\
\cmidrule(lr){2-3} \cmidrule(lr){4-5}
& Transductive & Inductive & Transductive & Inductive \\
\midrule
DyRep
& $0.7932 \pm 0.0143$ & $0.6310 \pm 0.0243$
& $0.7598 \pm 0.0236$ & $0.7245 \pm 0.0287$\\

JODIE
& $0.7876 \pm 0.0114$ & $0.7063 \pm 0.0192$
& $0.7346 \pm 0.0111$ & $0.7610 \pm 0.0141$\\

TGAT
& $0.7404 \pm 0.0088$ & $\underline{0.7416 \pm 0.0091}$
& $0.7269 \pm 0.0071$ & $0.7292 \pm 0.0065$ \\

TGN-no-mem
& $0.7404 \pm 0.0088$ & $\underline{0.7416 \pm 0.0091}$
& $0.7269 \pm 0.0071$ & $0.7292 \pm 0.0065$ \\

TGN-id
& $\underline{0.8417 \pm 0.0282}$ & $0.6499 \pm 0.0458$
& $0.7783 \pm 0.0330$ & $0.7066 \pm 0.0349$ \\

TGN-time
& $0.8398 \pm 0.0223$ & $0.6760 \pm 0.0303$
& $\underline{0.8001 \pm 0.0194}$ & $\underline{0.7801 \pm 0.0159}$\\

\textbf{TGN-SEAL}
& $\pmb{0.9288 \pm 0.0051}$ & $\pmb{0.9206 \pm 0.0064}$
& $\pmb{0.9249 \pm 0.0024}$ & $\pmb{0.9372 \pm 0.0027}$\\
\bottomrule
\end{tabular}%
}
\end{table*}

\begin{table*}[t]
\centering
\caption{Performance comparison on Email datasets (department 3 and 4).
\textbf{Bold}: best performance; \underline{Underline}: second best.}
\label{table:tgn-seal-result3}%
\resizebox{\textwidth}{!}{%
\begin{tabular}{lcccc}
\toprule
Method & \multicolumn{2}{c}{Email-3} & \multicolumn{2}{c}{Email-4} \\
\cmidrule(lr){2-3} \cmidrule(lr){4-5}
& Transductive & Inductive & Transductive & Inductive \\
\midrule
DyRep
& $0.6101 \pm 0.0075$ & $0.4936 \pm 0.0152$
& $0.7529 \pm 0.0150$ & $0.6956 \pm 0.0196$ \\

JODIE
& $0.6212 \pm 0.0076$ & $0.5260 \pm 0.0125$
& $0.7093 \pm 0.0338$ & $0.7211 \pm 0.0198$ \\

TGAT
& $\underline{0.6953 \pm 0.0042}$ & $\underline{0.7383 \pm 0.0050}$
& $0.7597 \pm 0.0054$ & $0.7162 \pm 0.0059$ \\

TGN-no-mem
& $0.6939 \pm 0.0052$ & $0.7342 \pm 0.0058$
& $0.7597 \pm 0.0054$ & $0.7162 \pm 0.0059$ \\

TGN-id
& $0.6215 \pm 0.0130$ & $0.5046 \pm 0.0079$
& $0.7633 \pm 0.0304$ & $0.6709 \pm 0.0401$ \\

TGN-time
& $0.6353 \pm 0.0054$ & $0.5139 \pm 0.0145$
& $\underline{0.7866 \pm 0.0156}$ & $\underline{0.7394 \pm 0.0194}$ \\

\textbf{TGN-SEAL}
& $\pmb{0.7741 \pm 0.0099}$ & $\pmb{0.7926 \pm 0.0097}$
& $\pmb{0.9007 \pm 0.0153}$ & $\pmb{0.8992 \pm 0.0229}$ \\
\bottomrule
\end{tabular}%
}
\end{table*}

\section{Discussion}

TGN-SEAL demonstrates that combining continuous-time memory with localized structural learning substantially improves dynamic link prediction. While standard TGNs model temporal interaction histories effectively, they often underperform in sparse or structurally heterogeneous networks where topology carries critical predictive signals. Incorporating enclosing subgraphs enables the model to capture higher-order dependencies and structural motifs that purely temporal approaches overlook, yielding consistent gains across datasets and evaluation settings.

Our results indicate that most predictive structural information is concentrated within local neighborhoods. In line with the $\gamma$-decaying heuristic theory, increasing the hop radius $k$ does not necessarily improve MAP, as the benefit depends on the specific graph topology. However, larger $k$ always increases computational cost. This observation highlights that small enclosing subgraphs are often sufficient for accurate link prediction while keeping the framework computationally efficient.

The use of DRNL labeling further enhances learning by encoding node roles relative to the target pair, allowing the GNN to distinguish structurally meaningful positions within enclosing subgraphs. In addition, aggregating neighbor memories mitigates the staleness problem of memory-based temporal models, enabling accurate predictions even for inactive nodes. The delayed memory update mechanism ensures strict temporal causality, preventing information leakage during training.

The primary limitation of TGN-SEAL lies in the computational overhead associated with subgraph extraction and processing, which increases per-epoch runtime. Although the model converges in fewer epochs, scalability may become challenging in extremely dense or high-frequency networks. Nevertheless, the tunable neighborhood radius provides a practical mechanism for balancing accuracy and efficiency.


\section*{Abbreviations}

\begin{description}
  \item[CDR] Call Data Record
  \item[CTDG] Continuous-Time Dynamic Graph
  \item[DRNL] Double-Radius Node Labeling
  \item[DTDG] Discrete-Time Dynamic Graph
  \item[GNN] Graph Neural Network
  \item[mAP] mean Average Precision
  \item[MLP] Multi-Layer Perceptron
  \item[PRM] Probabilistic Relational Model
  \item[SEAL] Subgraph Embeddings and Attributes for Link prediction
  \item[SVM] Support Vector Machine
  \item[TGN] Temporal Graph Network
  \item[WLNM] Weisfeiler-Lehman Neural Machine
  \item[WL] Weisfeiler-Lehman
\end{description}


\section*{Declarations}

\subsection*{Availability of data and materials}
The datasets analysed during the current study are available in the Reality Commons repository: \url{http://realitycommons.media.mit.edu/realitymining.html} and Stanford Large Network Dataset Collection: \url{https://snap.stanford.edu/data/}.  
The implementation code used in this study is openly available at: \url{https://github.com/nssajadi/tgn-seal/tree/main}.

\subsection*{Competing interests}
The authors declare that they have no competing interests.

\subsection*{Funding}
No funding was received for conducting this study.

\subsection*{Authors' contributions}
Nafiseh Sadat Sajadi conceived the research idea, designed and implemented the algorithms and experiments, analysed the results, and prepared the initial draft of the manuscript.  
Dr. Behnam Bahrak and Dr. Mahdi Jafari Siavoshani provided guidance on methodological and technical aspects, and contributed to the review and editing of the manuscript.  
All authors read and approved the final version of the manuscript.

\subsection*{Acknowledgements}
Not applicable.

\bibliography{sn-bibliography}

\end{document}